\documentclass[preprint,12pt]{elsarticle}
\usepackage{amsmath}
\usepackage{amssymb}
\usepackage[pdftex,colorlinks]{hyperref}
\usepackage{multirow}
\usepackage{enumerate}
\usepackage{cases}

\makeatletter

\newcommand {\Rmnum} [1] {\expandafter \@slowromancap \romannumeral #1@}
\makeatother
\newtheorem{Definition}{Definition}
\newtheorem{Protocol}{Protocol}

\biboptions{compress}

\begin{document}

\begin{frontmatter}

\title{Quantum oblivious transfer and bit commitment protocols based on two non-orthogonal states coding}


\author{Li Yang}
\address{State Key Laboratory of Information Security,
 Institute of Information Engineering, Chinese Academy of Sciences, Beijing 100093, China}

\begin{abstract}
Oblivious transfer protocols (R-OT and OT$_{1}^{2}$) are presented based on non-orthogonal states transmission, and the bit commitment protocols on the top of OT$_{1}^{2}$ are constructed. Although these OT protocols are all unconditional secure, the bit commitment protocols based on OT protocols are not secure against attack similar to that presented by no-go theorem. 
\end{abstract}


\end{frontmatter}


No-go theorem \cite{D97,HH97} declares that there is no quantum bit commitment protocol with unconditional security in both binding and concealing. Here we consider the construction of oblivious transfer and bit commitment protocols based on two non-orthogonal quantum states transmission, as that in \cite{LB06,L10}. Although the OT protocols and the construction from OT to BC are unconditionally secure, the BC protocols we finally have are not unconditionally secure ones. Then, we encounter a thought-provoking problem.
\section{Quantum random oblivious transfer}

\begin{Definition}
\emph{\textbf{(Random Oblivious Transfer (R-OT) Channel)}}

Alice sends a random bit $r$ to Bob via a channel, if
\begin{enumerate}
\item Bob obtains the bit value $r$ with a probability $p$ satisfies $0<\beta < p < \alpha$, $\alpha <\frac{1}{2}$, $\alpha$ and $\beta$ are any two real numbers;
\item Alice cannot know whether Bob has get the value of her bit.
\end{enumerate}
Then, the channel is named a R-OT channel (an extended Rabin's OT channel).
\end{Definition}
~~~~Let two quantum states $|\Psi_{0}\rangle$, $|\Psi_{1}\rangle$ satisfy $\langle\Psi_{0}|\Psi_{1}\rangle=\cos\theta$, we can distinguish them with probability $p=1-\cos\theta$, by some POVM process, which is the optimal result. Here after we choose $\theta=\frac{\pi}{4}$, then the optimal scheme to distinguish $|\Psi_{0}\rangle$ and $|\Psi_{1}\rangle$ will be of a success probability $p=1-\frac{\sqrt{2}}{2}$. To simplify the protocol to be a practical one, we choose an easy way of measurement: the receiver Bob randomly chooses one of the two bases, $B_{0}=\{|\Psi_{0}\rangle, |\Psi_{0}\rangle^{\perp}\}$ and $B_{1}=\{|\Psi_{1}\rangle, |\Psi_{1}\rangle^{\perp}\}$, as his measurement bases. When his measurement results in $|\Psi_{x}\rangle^{\perp}$, Bob admits $|\Psi_{x\oplus 1}\rangle$ as that comes from Alice; otherwise, he concludes no result.

In this case, Bob's probability of conclusive bits is $\frac{1}{2}\cos^{2}\frac{\pi}{4}=\frac{1}{4}$. We put forward a quantum R-OT protocol as follows:
\begin{Protocol}
\emph{\textbf{Quantum R-OT}}
\begin{enumerate}
\item Alice generates random bit string $(r_{1},\ldots,r_{n})$, and sends $|\Psi_{r_{1}}\rangle, \ldots, |\Psi_{r_{n}}\rangle$ to Bob.
\item Bob chooses $B_{1}$ or $B_{2}$ randomly to measure the qubit string coming from Alice. He accepts a bit as a conclusive bit if and only if his measurement results in state $|\Psi_{x}\rangle^{\perp}$, and takes the value of this bit as $x\oplus 1$.
\end{enumerate}
\end{Protocol}

It can be seen that an honest Bob will get about $\frac{1}{4}n$ conclusive bits, though a malicious Bob can get up to $(1-\frac{\sqrt{2}}{2})n$ conclusive bits with general individual attack. If Alice can get a bit's value and ensure that it is a conclusive bit, the qubit Bob obtained must be in a pure state. Therefore, Alice cannot execute EPR attack, and then, she cannot know whether a bit with a given value has been taken as a conclusive bit by Bob.

\section{One-out-of-two oblivious transfer(OT$_{1}^{2}$)}
Let us construct an OT$_{1}^{2}$ protocol based on protocol 1. The construction of OT$_{1}^{2}$ on top of R-OT is proposed firstly by Cr\'{e}peau \cite{Crpeau88} in 1988. For parameter $\alpha=\frac{1}{16}$, we have $k\equiv \lfloor(\frac{1}{4}-\alpha)n\rfloor=\lfloor\frac{3}{16}n\rfloor$, $\lfloor\frac{3}{8}n\rfloor-1\leq2k\leq\lfloor\frac{3}{8}n\rfloor$. Suppose the probability of the cases that the number of conclusive bits obtained by an honest Bob is more than $k$ be $p_{1}$, and the probability of the cases that the number of conclusive bits obtained by a malicious Bob is equal to or greater than $2k$ be $p_{2}$.
\begin{Protocol}\textbf{\emph{(OT$_{1}^{2}$)}}
~~~~~~~~~~~~~~~~~~~~~
\begin{enumerate}
\item After executing protocol 1 with Alice, Bob chooses $k$ bits $r_{i_{1}},\ldots, r_{i_{k}}$ from his conclusive bits, and define $I\equiv\{i_{1},\ldots, i_{k}\}$; then he choose randomly $k$ bits $r_{j_{1}},\ldots, r_{j_{k}}$ from set $\{1,\ldots,n\}\backslash I$, and define $J\equiv \{j_{1},\ldots, j_{k}\}$.
\item Bob sends $\{X, Y\}$ to Alice, chooses randomly $\{X, Y\}= \{ I, J\}$ or $\{X, Y\}= \{ J, I\}$ according to his random bit with value $m$.
\item After receiving $(X, Y)$, Alice computes:
\begin{eqnarray*}\label{eq:3}
{s_0} = \mathop  \oplus \limits_{i \in X} {r_i},{s_1} = \mathop  \oplus \limits_{i \in Y} {r_i}
\end{eqnarray*}
then encrypts her messages $b_0$ and $b_1$ as
\begin{eqnarray*}\label{eq:4}
{c_0} = {b_0} \oplus {s_0},{c_1} = {b_1} \oplus {s_1},
\end{eqnarray*}
and sends them to Bob.
\item Bob computes $s=\bigoplus_{i \in I}r_i$, and decrypts that coming from Alice to obtain either $b_0$ or $b_1$.
\end{enumerate}
\end{Protocol}
It is obvious that the above protocol is a secure OT$_{1}^{2}$ protocol if and only if the following conditions are satisfied:
\begin{enumerate}
\item $p_{1}\rightarrow1$ exponentially  as $n\rightarrow\infty$.
\item $p_{2}\rightarrow0$ exponentially  as $n\rightarrow\infty$.
\end{enumerate}
We can easily find that these conditions really holds as $\alpha=\frac{1}{16}$.

\section{Bit commitment protocol}
The entanglement state is needed in the attack on quantum bit commitment protocol according to no-go theorem. Consider of the famous averment of resisting entanglement with entanglement, as used in the construction of quantum error-correcting codes, we present a bit commitment protocol as follows:
~\\
~\\

\begin{Protocol}\textbf{\emph{(Bit commitment)}}
~~~~~~~~~~~~~~~~~~~~~~~~~~~~~~~~~~~~~~~~~~~~~~~~~~~~~~~~~~~~~~~~~~~~~~~~~~~~~~~~~~~~~~~~~~~~~~~~~~~~~~~~~
~~~~~~~~~~~~~~~~~~~~~~~~~~~~~~~~~

Commit phase:
\begin{enumerate}
\item Bob prepares quantum state $|\Phi_1^-\rangle,...,|\Phi_n^-\rangle$,
\begin{eqnarray}\label{phi}
|\Phi^-\rangle=\frac{1}{\sqrt{2}}(|0\rangle_{\Rmnum{1}}|0\rangle_{\Rmnum{2}}-|1\rangle_{\Rmnum{1}}|1\rangle_{\Rmnum{2}}).
\end{eqnarray}
and sends each first qubit of every pair to Alice.
\item Alice generates random bit string $(r_{1},\ldots,r_{n})$. When $r_i=0$, she keeps the ith qubit unchanged and sends it back to Bob; when $r_i=1$, she rotates the state along y axis (not Hadamard transformation) with transformation $U=R_y(-\frac{\pi}{2})$, where $R_y(-\frac{\pi}{2})=e^{i\frac{\pi}{4}Y}$, and sends the qubit back to Bob, that is
\begin{equation}\nonumber
\left\{
\begin{aligned}
&r_i=0,~~ |\Phi^-\rangle\longrightarrow |\Phi^-\rangle,\\
&r_i=1,~~ |\Phi^-\rangle\longrightarrow \frac{1}{\sqrt{2}}(|\Phi^-\rangle+|\Psi^+\rangle)
\end{aligned}
\right.
\end{equation}
since
\begin{equation}\nonumber
\begin{aligned}
&\frac{1}{\sqrt{2}}(e^{i\frac{\pi}{4}Y}|0\rangle_{\Rmnum{1}}\otimes|0\rangle_{\Rmnum{2}}-
e^{i\frac{\pi}{4}Y}|1\rangle_{\Rmnum{1}}\otimes|1\rangle_{\Rmnum{2}})\\
=&\frac{1}{\sqrt{2}}(|+\rangle_{\Rmnum{1}}|0\rangle_{\Rmnum{2}}+|-\rangle_{\Rmnum{1}}|1\rangle_{\Rmnum{2}})\\
=&\frac{1}{2}\left[(|0\rangle_{\Rmnum{1}}|0\rangle_{\Rmnum{2}}-|1\rangle_{\Rmnum{1}}|1\rangle_{\Rmnum{2}})
+(|1\rangle_{\Rmnum{1}}|0\rangle_{\Rmnum{2}}+|0\rangle_{\Rmnum{1}}|1\rangle_{\Rmnum{2}})\right]\\
=&\frac{1}{\sqrt{2}}(|\Phi^-\rangle+|\Psi^+\rangle).
\end{aligned}
\end{equation}
\item Bob chooses $B_{0}$ or $B_{1}$ randomly to measure the qubit string coming from Alice, bases
\begin{footnotesize}
\begin{equation}\nonumber
\begin{aligned}
&B_0=\left\{|\Phi^-\rangle, |\Phi^+\rangle, |\Psi^-\rangle, |\Psi^+\rangle\right\}\\
&B_1=\left\{\frac{1}{\sqrt{2}}(|\Phi^-\rangle+|\Psi^+\rangle), \frac{1}{\sqrt{2}}(|\Phi^-\rangle-|\Psi^+\rangle),
\frac{1}{\sqrt{2}}(|\Phi^+\rangle+|\Psi^-\rangle),
\frac{1}{\sqrt{2}}(|\Phi^+\rangle+|\Psi^-\rangle)\right\}
\end{aligned}
\end{equation}
\end{footnotesize}
   He accepts a bit as a conclusive bit if and only if his measurement results in states $|\Psi^+\rangle$ or $\frac{1}{\sqrt{2}}(|\Phi^-\rangle-|\Psi^+\rangle)$. When it results in $|\Psi^+\rangle$, he takes the value of this bit as $1$; when it results in $\frac{1}{\sqrt{2}}(|\Phi^-\rangle-|\Psi^+\rangle)$, he takes the value of this bit as $0$. It can be seen the probability of a conclusive bit is $\frac{1}{4}$.

\item Bob chooses $k$ bits $r_{i_{1}},\ldots, r_{i_{k}}$ from his conclusive bits, where $k=\frac{3}{16}n$ and define $I\equiv\{i_{1},\ldots, i_{k}\}$; then he chooses $B_{1}$ or $B_{2}$ randomly to measure the qubit string coming from Alice. He choose randomly $k$ bits $r_{j_{1}},\ldots, r_{j_{k}}$ from set $\{1,\ldots,n\}\backslash I$, and define $J\equiv \{j_{1},\ldots, j_{k}\}$. He sends $\{X, Y\}$ to Alice, chooses randomly $\{X, Y\}= \{ I, J\}$ or $\{X, Y\}= \{ J, I\}$ according to his random bit with value $m$.

\item After receiving $(X, Y)$, Alice computes:
\begin{eqnarray*}\label{eq:3}
{s_0} = \mathop  \oplus \limits_{i \in X} {r_i},{s_1} = \mathop  \oplus \limits_{i \in Y} {r_i}
\end{eqnarray*}
then encrypts her messages $b_0$ and $b_1$ as
\begin{eqnarray*}\label{eq:4}
{c_0} = {b_0} \oplus {s_0},{c_1} = {b_1} \oplus {s_1},
\end{eqnarray*}
and sends them to Bob.

\item Bob computes $s=\bigoplus_{i \in I}r_i$, and decrypts that coming from Alice to obtain either $b_0$ or $b_1$.

\item Alice divides randomly her commit value as $b= b_{0}^{(i)}\oplus b_{1}^{(i)}$, $i= 1,\ldots, l$ and executes Step $1-6$ with Bob $l$ times for sending $\{b_0^{(i)},b_1^{(i)}|i=1,...,l\}$.
\end{enumerate}
~~~~~Open phase:
\begin{enumerate}

\item Alice opens $\{b_{0}^{(i)}, b_{1}^{(i)}; r_{i_{1}(i)}^{(i)},\ldots,r_{i_{k}(i)}^{(i)}; r_{j_{1}(i)}^{(i)},\ldots,r_{j_{k}(i)}^{(i)}|i= 1,\ldots l\}$..

\item Bob verifies whether $\{b_{0}^{(i)}, b_{1}^{(i)}; r_{i_{1}(i)}^{(i)},\ldots,r_{i_{k}(i)}^{(i)}; r_{j_{1}(i)}^{(i)},\ldots,r_{j_{k}(i)}^{(i)}|i= 1,\ldots l\}$ is consistent with his $\{b_{m_{i}}^{(i)}; r_{i_{1}(i)}^{(i)},\ldots,r_{i_{k}(i)}^{(i)}|i= 1,\ldots l\}$ and those conclusive bits in J. If the consistency holds, he admits Alice's commit value as $b$.
\end{enumerate}
\end{Protocol}

\section{Analysis on security}
\subsection{The security of bit commitment protocol based on Protocol 2}
From Protocol 2, Bob can obtain either $b_0$ or $b_1$. Alice divides randomly her commit value as $b= b_{0}^{(i)}\oplus b_{1}^{(i)}$, $i= 1,\ldots, l$ and executes Protocol 2 with Bob $l$ times for sending $\{b_0^{(i)},b_1^{(i)}|i=1,...,l\}$, then a bit commitment protocol is constructed.

This bit commitment protocol can be attacked according to no-go theorem. When Alice commits $0$ or $1$, she prepares
\begin{eqnarray}
\begin{aligned}\label{1}
|0\rangle=\sum_{i=1}^{2k} \alpha_i|e_i\rangle_A\otimes |\Psi_{r_i}\rangle_B,\\
|1\rangle=\sum_{j=1}^{2k} \beta_j|e'_j\rangle_A\otimes |\Psi_{r_j}\rangle_B,
\end{aligned}
\end{eqnarray}
respectively, where
\begin{eqnarray}
\begin{aligned}\label{2}
&\bigoplus_{i=1}^{2k}r_i=c_0\oplus c_1,\\
&\bigoplus_{j=1}^{2k}r_j=c_0\oplus c_1 \oplus1,
\end{aligned}
\end{eqnarray}
Define $|\phi_0\rangle$, $|\phi_1\rangle$ are two pure states on the composite system, and $\rho_0^B=Tr(|\phi_0\rangle\langle\phi_0|)$, $\rho_1^B=Tr(|\phi_1\rangle\langle\phi_1|)$. The fidelity of the two density matrices is
\begin{eqnarray}
F(\rho_0^B, \rho_1^B)=max|\langle\phi_0|\phi_1\rangle|
\end{eqnarray}
From Equation (\ref{1}) (\ref{2}), it can be seen that Bob can distinguish the two density matrices with a small probability.
\begin{eqnarray}
F(\rho_0^B, \rho_1^B)=1-\delta
\end{eqnarray}
where $\delta>0$ is small. If Alice wants to change the value of commitment $0$ to $1$, she can apply a unitary transformation acting on $A$ alone to obtain $|\phi_0\rangle$, which satisfies
\begin{eqnarray}
|\langle\phi_0|1\rangle|=F(\rho_0^B, \rho_1^B)=1-\delta
\end{eqnarray}
As $|\phi_0\rangle$ and $|1\rangle$ are so similar, Bob hardly can detect the cheating Alice.

\subsection{The security of bit commitment protocol based on a modified protocol of Protocol 2}
We modify Step $3$ and Step $4$ of Protocol 2 as follows:

\begin{enumerate}
\item After receiving $(X, Y)$, Alice encrypts her messages $b_{0}$ and $b_{1}$ as
\begin{equation}\nonumber
\left\{
\begin{array}{c}
c_{0}=E_{(r_{x_1},...,r_{x_k})}(R_{0},b_{0})\in \{0,1\}^{k+1},\{x_1,...x_k\}=X,\\
c_{1}=E_{(r_{y_1},...,r_{y_k})}(R_{1},b_{1})\in \{0,1\}^{k+1},\{y_1,...y_k\}=Y,
\end{array}
\right.
\end{equation}
 where $x,y$ are the encryption keys, $R_{0},R_{1}\in \{0,1\}^{k}$ are Alice's local random bit-string. Alice sends $c_0$, $c_1$ to Bob and keeps $R_0$, $R_1$ secret.

\item Bob decrypts that coming from Alice $D_{key}(c_m)$
 to obtain $(R_{m}, b_m)=D_{key}(c_{m})$ with $key=(r_{i_1},...,r_{i_k})$.
\end{enumerate}
Alice divides randomly her commit value as $b= b_{0}^{(i)}\oplus b_{1}^{(i)}$, $i= 1,\ldots, l$ and executes the $OT_1^2$ with Bob $l$ times for sending $\{b_0^{(i)},b_1^{(i)}|i=1,...,l\}$, then another bit commitment protocol is constructed. In open phase of the bit commitment, the value of $R_0$ and $R_1$ should be opened to limit a cheating Alice.

Then we analyze whether $R_0$ and $R_1$ can limit Alice to apply a unitary transformation and not to be detected by Bob.
Suppose
\begin{eqnarray}
\begin{aligned}
&(Q_0(r_{x_1},...,r_{x_k})\oplus Q_1(r_{y_1},...,r_{y_k}), d_0(r_{x_1},...,r_{x_k})\oplus d_1(r_{y_1},...,r_{y_k}))\\
=&(D_{(r_{x_1},...,r_{x_k})}(R_0)\oplus D_{(r_{y_1},...,r_{y_k})}(R_1), D_{(r_{x_1},...,r_{x_k})}(b_0)\oplus D_{(r_{y_1},...,r_{y_k})}(b_1))\\
=&(R_0\oplus R_1, b_0\oplus b_1).
\end{aligned}
\end{eqnarray}
 Generally, if Alice tries to apply a unitary transformation acting on A, the change of $d_0(r_{x_1},...,r_{x_k})\oplus d_1(r_{y_1},...,r_{y_k})$ could not be detected by Bob. Assume $\{X,Y\}=\{I,J\}$ and Bob does not know the value of $\{r_{y_i},...,r_{y_k}\}$. As long as Alice applies this unitary transformation, she can get the value of $\{r_{x_1},...,r_{x_k},r_{y_1},...,r_{y_{(i-1)}},r_{y'_i},...,r_{y'_k}\}$, which allows Alice computes
 \begin{eqnarray}
 Q_0(r_{x_1},...,r_{x_k})   \oplus Q_1(r_{y_1},...,r_{y_{(i-1)}},r_{y'_i},...,r_{y'_k})=(R_0\oplus R'_1).
 \end{eqnarray}
 The value of $R_0$ is consistent with that Bob obtains and the change of $R_1$ can not be detected by Bob. Alice can always cheat successfully.

It is interesting that the quantum OT$_1^2$ protocols are secure but the bit commitment based on them can be attacked according to no-go theorem. However, in classical cryptography, the bit commitment protocol on the top of a secure OT$_1^2$ protocol is secure if only the construction is secure. This interesting results remaind us that the composable security of quantum protocols is worth to be investigated theoretically.

\subsection{The security of Protocol 3}
As Alice attacks according to no-go theorem, she may make her probe qubit entangled with that comes from Bob. Suppose the state has been changed to $\frac{1}{\sqrt{2}}(|0\rangle_{A}|0\rangle_{B\Rmnum{1}}|0\rangle_{B\Rmnum{2}}-
|1\rangle_{A}|1\rangle_{B\Rmnum{1}}|1\rangle_{B\Rmnum{2}})$.
For $r_i=1$, the state becomes

\begin{eqnarray}\label{eq:jiu}
\begin{aligned}
&\frac{1}{\sqrt{2}}(|0\rangle_A\otimes e^{i\frac{\pi}{4}y}|0\rangle_{B\Rmnum{1}}\otimes|0\rangle_{B\Rmnum{2}}-
|1\rangle_A e^{i\frac{\pi}{4}y}|1\rangle_{B\Rmnum{1}}\otimes|1\rangle_{B\Rmnum{2}})\\
=&\frac{1}{\sqrt{2}}(|0\rangle_{A}|+\rangle_{B\Rmnum{1}}|0\rangle_{B\Rmnum{2}}
+|1\rangle_{A}|-\rangle_{B\Rmnum{1}}|1\rangle_{B\Rmnum{2}})\\
=&\frac{1}{2}\left(|0\rangle_{A}|0\rangle_{B\Rmnum{1}}|0\rangle_{B\Rmnum{2}}+
|0\rangle_{A}|1\rangle_{B\Rmnum{1}}|0\rangle_{B\Rmnum{2}}+
|1\rangle_{A}|0\rangle_{B\Rmnum{1}}|1\rangle_{B\Rmnum{2}}-
|1\rangle_{A}|1\rangle_{B\Rmnum{1}}|1\rangle_{B\Rmnum{2}}\right)\\
=&\frac{1}{2}\left(|+\rangle_{A}|\Phi^-\rangle_{B}+
|-\rangle_{A}|\Phi^+\rangle_{B}+
|+\rangle_{A}|\Psi^+\rangle_{B}+
|-\rangle_{A}|\Psi^-\rangle_{B}\right),
\end{aligned}
\end{eqnarray}
For $r_i=0$, the state becomes
\begin{eqnarray}
\begin{aligned}\label{eq:9}
&\frac{1}{\sqrt{2}}(|0\rangle_{\Rmnum{3}}|0\rangle_{\Rmnum{1}}|0\rangle_{\Rmnum{2}}-
|1\rangle_{\Rmnum{3}}|1\rangle_{\Rmnum{1}}|1\rangle_{\Rmnum{2}})\\
=&\frac{1}{2}\left[|0\rangle_{\Rmnum{3}}(|\Phi^-\rangle_{\Rmnum{1},\Rmnum{2}}+
|\Phi^+\rangle_{\Rmnum{1},\Rmnum{2}})-
|1\rangle_{\Rmnum{3}}(|\Phi^+\rangle_{\Rmnum{1},\Rmnum{2}}-
|\Phi^-\rangle_{\Rmnum{1},\Rmnum{2}})\right]
\end{aligned}
\end{eqnarray}

Suppose Bob randomly chooses $B_0$ and $B_1$ to measure the two states. When he chooses $B_0$ to measure the state in the end of Equation (9), the probability that the measurement results in $|\Phi^-\rangle$, $|\Phi^+\rangle$, $|\Psi^-\rangle$, $|\Psi^+\rangle$ is $\frac{1}{4}$, respectively. Since $|\Phi^+\rangle$, $|\Psi^-\rangle$ is the result that should not exist without Alice's attack, Bob can recognize Alice's attack with a probability $\frac{1}{2}$. Similarly, the probability of recognizing the attack of Alice in the other three cases is $\frac{1}{2}$. Since there is $n$ qubits transmitted, the probability of Alice's successful attack is $\frac{1}{2^n}$. Thus, this no-go theorem type attack can not work.

However, there is a better attack. The entangled state prepared by Bob can be regarded as $|\psi_B\rangle=\sum_{i,j}\alpha_{ij}|\phi_i\rangle_{B\Rmnum{1}}|\phi_j\rangle_{B\Rmnum{2}}$. Then he sends the first register to Alice. Alice dose a controlled unitary transformation instead of the protocol operation.
\begin{align}\label{eq:9}
|\psi\rangle=&\frac{1}{n}\sum_{r_k}|r_1,...,r_n\rangle_A U_{B\Rmnum{1}}(r_1,...,r_n)\bigotimes_{i=1}^n |\Phi_i^-\rangle_B \notag\\
=&\frac{1}{n}\sum_{r_k}|r_1,...,r_n\rangle_A U_{B\Rmnum{1}}(r_1,...,r_n)\sum_{i,j}\alpha_{ij}|\phi_i\rangle_{B\Rmnum{1}}|\phi_j\rangle_{B\Rmnum{2}}\notag\\
=&\frac{1}{n}\sum_{r_k,i,j}\alpha_{ij}|r_1,...,r_n\rangle_A \left[U_{B\Rmnum{1}}(r_1)\otimes...\otimes U_{B\Rmnum{1}}(r_n)\right] |\phi_i\rangle_{B\Rmnum{1}} |\phi_j\rangle_{B\Rmnum{2}}\notag \\
=&\frac{1}{n}\sum_{r_k,i,j}\alpha_{ij}|r_1,...,r_n\rangle_A \left[U_{B\Rmnum{1}}(r_1)|\phi_i^1\rangle_{B\Rmnum{1}}\right] \otimes...\otimes \left[U_{B\Rmnum{1}}(r_n)|\phi_i^n\rangle_{B\Rmnum{1}}\right]|\phi_j\rangle_{B\Rmnum{2}} \notag\\
=&\frac{1}{n}\sum_{r_k,i,j}\alpha_{ij}|r_1,...,r_n\rangle_A |\phi_i^{1'}\rangle_{B\Rmnum{1}} \otimes...\otimes |\phi_i^{n'}\rangle_{B\Rmnum{1}}|\phi_j\rangle_{B\Rmnum{2}} \notag\\
=&\sum_{r_k,i,j}\alpha'_{ijk}|r_k\rangle_A \otimes |\phi'_i\rangle_{B\Rmnum{1}}\otimes |\phi_j\rangle_{B\Rmnum{2}}
\end{align}
It is easy to see that the control qubits in Alice's hands are entangled with Bob's registers. When Alice commits $0$, the whole state is
 \begin{eqnarray}
 |0\rangle=\sum_{i,j,r_v}\alpha_{ijv}|r_v\rangle_A \otimes |\phi'_i\rangle_{B\Rmnum{1}}\otimes |\phi_j\rangle_{B\Rmnum{2}},
 \end{eqnarray}
where $\bigoplus_{v=1}^{2k}r_v=c_0\oplus c_1$. When Alice commits $1$, the whole state is
 \begin{eqnarray}
 |1\rangle=\sum_{i,j,r_u}\alpha_{iju}|r_u\rangle_A \otimes |\phi'_i\rangle_{B\Rmnum{1}}\otimes |\phi_j\rangle_{B\Rmnum{2}},
 \end{eqnarray}
where $\bigoplus_{u=1}^{2k}r_u=c_0\oplus c_1 \oplus 1$.

Imagine a system A attached to Bob's system B\Rmnum{1} and B\Rmnum{2}. There are many pure states $|\psi_0\rangle$ and $|\psi_1\rangle$ on the composite system such that
$\rho_0^B=Tr_A(|\psi_0\rangle\langle\psi_0|)$, $\rho_1^B=Tr_A(|\psi_1\rangle\langle\psi_1|)$. Bob can distinguish the two density matrices with a small probability.
\begin{eqnarray}
F(\rho_0^B, \rho_1^B)=1-\delta
\end{eqnarray}
where $\delta>0$ is small. If Alice wants to change the value of commitment $0$ to $1$, she can apply a unitary transformation acting on $A$ alone to obtain $|\psi_0\rangle$, which satisfies
\begin{eqnarray}
|\langle\psi_0|1\rangle|=F(\rho_0^B, \rho_1^B)=1-\delta
\end{eqnarray}
As $|\psi_0\rangle$ and $|1\rangle$ are so similar, Bob hardly can detect the cheating Alice.

\section{Improvement of the bit commitment}
\subsection{Improvement of Protocol 3}
In Step $1$ of the bit commitment, Bob rotates the first qubit along y axis with a random degree $\alpha$ before sending it to Alice, that is
\begin{equation}\nonumber
\left\{
\begin{aligned}
&|0\rangle_{\Rmnum{1}}\longrightarrow |\alpha\rangle_{\Rmnum{1}},\\
&|1\rangle_{\Rmnum{1}}\longrightarrow |\alpha+\frac{\pi}{2}\rangle_{\Rmnum{1}}
\end{aligned}
\right.
\end{equation}

Since both Bob's rotation in Step $1$ and Alice's rotation in Step $2$ are along y axis, the two unitary transformations are exchangeable. For Bob, he rotates the qubit coming from Alice along y axis with $-\alpha$ and executes the remaining part of the bit commitment. The correctness of the protocol can also be ensured. For Alice, the attack mentioned in Section 4.3 also works.

\subsection{A more simple bit commitment protocol}
Here is a more simple and practical bit commitment protocol.
\begin{Protocol}
~~~~~~~~~~~~~~~~~~~~~~~~~~~~~~~~~~~~~~~~~~~~~~~~~~~~~~~~~~~~~~~~~~~~~~~~~~~~~~~~~~~~~~~~~~~~~~~~~~~~~~~~~
~~~~~~~~~~~~~~~~~~~~~~~~~~~~~~~~~

Commit phase:
\begin{enumerate}
\item Bob prepares a random qubit string $|\alpha_1\rangle,...,|\alpha_n\rangle$ and sends it to Alice. Each secret random value $\alpha_i$ represents the angle between the state $|\alpha_i\rangle$ and the state $|0\rangle$ along y axis.

\item Alice generates random bit string $(r_{1},\ldots,r_{n})$. When $r_i=0$, she keeps the ith qubit unchanged and sends it back to Bob; when $r_i=1$, she rotates the state along y axis (not Hadamard transformation) with transformation $U=R_y(-\frac{\pi}{2})$, where $R_y(-\frac{\pi}{2})=e^{i\frac{\pi}{4}Y}$, and sends the qubit back to Bob, that is
\begin{equation}\nonumber
\left\{
\begin{aligned}
&r_i=0,~~ |\alpha\rangle \longrightarrow |\alpha\rangle,\\
&r_i=1,~~ |\alpha\rangle \longrightarrow |\alpha+\frac{\pi}{4}\rangle
\end{aligned}
\right.
\end{equation}

\item For each qubit coming from Alice, Bob rotates the ith qubit along y axis with $-\alpha_i$ and chooses $B_{0}$ or $B_{1}$ randomly to measure, bases
\begin{equation}\nonumber
\begin{aligned}
&B_0=\left\{|0\rangle, |1\rangle\right\}\\
&B_1=\left\{|+\rangle, |-\rangle\right\}
\end{aligned}
\end{equation}
   He accepts a bit as a conclusive bit if and only if his measurement results in states $|1\rangle$ or $|-\rangle$. When it results in $|1\rangle$, he takes the value of this bit as $1$; when it results in $|-\rangle$, he takes the value of this bit as $0$. It can be seen the probability of a conclusive bit is $\frac{1}{4}$.

\item Bob chooses $k$ bits $r_{i_{1}},\ldots, r_{i_{k}}$ from his conclusive bits, where $k=\frac{3}{16}n$ and define $I\equiv\{i_{1},\ldots, i_{k}\}$; then he chooses $B_{1}$ or $B_{2}$ randomly to measure the qubit string coming from Alice. He choose randomly $k$ bits $r_{j_{1}},\ldots, r_{j_{k}}$ from set $\{1,\ldots,n\}\backslash I$, and define $J\equiv \{j_{1},\ldots, j_{k}\}$. He sends $\{X, Y\}$ to Alice, chooses randomly $\{X, Y\}= \{ I, J\}$ or $\{X, Y\}= \{ J, I\}$ according to his random bit with value $m$.

\item After receiving $(X, Y)$, Alice computes:
\begin{eqnarray*}\label{eq:3}
{s_0} = \mathop  \oplus \limits_{i \in X} {r_i},{s_1} = \mathop  \oplus \limits_{i \in Y} {r_i}
\end{eqnarray*}
then encrypts her messages $b_0$ and $b_1$ as
\begin{eqnarray*}\label{eq:4}
{c_0} = {b_0} \oplus {s_0},{c_1} = {b_1} \oplus {s_1},
\end{eqnarray*}
and sends them to Bob.

\item Bob computes $s=\bigoplus_{i \in I}r_i$, and decrypts that coming from Alice to obtain either $b_0$ or $b_1$.

\item Alice divides randomly her commit value as $b= b_{0}^{(i)}\oplus b_{1}^{(i)}$, $i= 1,\ldots, l$ and executes Step $1-6$ with Bob $l$ times for sending $\{b_0^{(i)},b_1^{(i)}|i=1,...,l\}$.
\end{enumerate}
~~~~~Open phase:
\begin{enumerate}

\item Alice opens $\{b_{0}^{(i)}, b_{1}^{(i)}; r_{i_{1}(i)}^{(i)},\ldots,r_{i_{k}(i)}^{(i)}; r_{j_{1}(i)}^{(i)},\ldots,r_{j_{k}(i)}^{(i)}|i= 1,\ldots l\}$..

\item Bob verifies whether $\{b_{0}^{(i)}, b_{1}^{(i)}; r_{i_{1}(i)}^{(i)},\ldots,r_{i_{k}(i)}^{(i)}; r_{j_{1}(i)}^{(i)},\ldots,r_{j_{k}(i)}^{(i)}|i= 1,\ldots l\}$ is consistent with his $\{b_{m_{i}}^{(i)}; r_{i_{1}(i)}^{(i)},\ldots,r_{i_{k}(i)}^{(i)}|i= 1,\ldots l\}$ and those conclusive bits in J. If the consistency holds, he admits Alice's commit value as $b$.
\end{enumerate}
\end{Protocol}

Suppose the committer has an ideal single-photon source, the quantum channel is a perfect channel without loss or error and the receiver has two perfect single-photon detectors, there is a more simple bit commitment protocol constructed directly as follows:
\begin{Protocol}
~~~~~~~~~~~~~~~~~~~~~~~~~~~~~~~~~~~~~~~~~~~~~~~~~~~~~~~~~~~~~~~~~~~~~~~~~~~~~~~~~~~~~~~~~~~~~~~~~~~~~~~~~
~~~~~~~~~~~~~~~~~~~~~~~~~~~~~~~~~

Commit phase:
\begin{enumerate}
\item Bob prepares a random qubit string $|\alpha^{(1)}_1\rangle,...,|\alpha^{(1)}_n\rangle,...... ,|\alpha^{(m)}_1\rangle,...,|\alpha^{(m)}_n\rangle$ and sends it to Alice. Each secret random value $\alpha^{(i)}_j$ represents the angle between the state $|\alpha^{(i)}_j\rangle$ and the state $|0\rangle$ along y axis.

\item Alice chooses $r^{(i)}\in\{0,1\}^{n}$ randomly, here $i=1,2,...,m$, $r^{(i)}=(r^{(i)}_1,r^{(i)}_2,$ $...,r^{(i)}_n)$ satisfies $F(r^{(i)})=b$, where $F(\cdot)$ is an $n^{th}_0$-order correlation immune Boolean function. When $r^{(i)}_j=0$, she keeps the $(n\times j-n+i)_{th}$ qubit unchanged and sends it back to Bob; when $r^{(i)}_j=1$, she rotates the state along y axis (not Hadamard transformation) with transformation $U=R_y(-\frac{\pi}{2})$, where $R_y(-\frac{\pi}{2})=e^{i\frac{\pi}{4}Y}$, and sends the qubit back to Bob as a piece of evidence for her commitment, that is
\begin{equation}\nonumber
\left\{
\begin{aligned}
&r^{(i)}_j=0,~~ |\alpha\rangle \longrightarrow |\alpha\rangle,\\
&r^{(i)}_j=1,~~ |\alpha\rangle \longrightarrow |\alpha+\frac{\pi}{4}\rangle
\end{aligned}
\right.
\end{equation}
\end{enumerate}
~~~~~Open phase:
\begin{enumerate}

\item Alice opens by declaring $b$ and the values of $r^{(i)}$.

\item Bob verifies by corresponding projective measurement: if $r^{(i)}_j=0$, Bob rotates the $(n\times j-n+i)_{th}$ qubit along y axis with $-\alpha_i$ and chooses $B_{0}$ or $B_{1}$ randomly to measure, bases
\begin{equation}\nonumber
\begin{aligned}
&B_0=\left\{|0\rangle, |1\rangle\right\}\\
&B_1=\left\{|+\rangle, |-\rangle\right\}
\end{aligned}
\end{equation}
Unless each results is matched, Bob has to break off the scheme.
\item Bob checks commitment value $b$. If $r^{(i)}$ satisfies $b=F(r^{(i)})$ for every $i$, Bob accepts the commitment value.
\end{enumerate}
\end{Protocol}

Quantum memory is not necessary in Protocol $5$ if Bob's measurement is executed in commit phase. Unless Bob uses two perfect single-photon detectors, Alice can attack the binding of the protocol by omitting one qubit in each $n$-qubit-string. This is the reason that we can not develop this protocol to be a practical one.

\section{Discussion}

Though these protocols are insecure theoretically, the attacks can hardly be applied in practice for two reasons: 1) Such as $2k=200$, the entry number of matrix $U_A$ is $2^{200}\times2^{200}$, which is greater than the number of atoms of the earth (approximately $10^{50}$). It means Alice cannot get the matrix actually. 2) The storage time of quantum state is limited. The bit commitment protocol can be executed after a period of time to prevent Alice executing her local unitary transformation with the quantum states in her hands. Define the protocols which cannot attack by these reasons are physically secure.

In this paper, we present  R-OT and OT$_{1}^{2}$ protocols and several bit commitment protocols on the top of OT$_{1}^{2}$.  The quantum bit commitment protocols we proposed are not beyond no-go theorem, but they are physically secure.
~\\
~\\

I would like to thank Ya-Qi Song, Guang-Ping He, Chong Xiang and Hai-Xia Xu for useful discussions. This work was supported by the National Natural Science Foundation of China under Grant No.61173157.

\end{document}